# How many candidates are needed to make elections hard to manipulate?

Vincent Conitzer*

Carnegie Mellon University

Computer Science Dept.

5000 Forbes Avenue

Pittsburgh, PA 15213, USA

conitzer@cs.cmu.edu

Jérôme Lang†

IRIT

Université Paul Sabatier

31062 Toulouse Cedex

France

lang@irit.fr

Tuomas Sandholm

Carnegie Mellon University

Computer Science Dept.

5000 Forbes Avenue

Pittsburgh, PA 15213, USA

sandholm@cs.cmu.edu

**Abstract**

In multiagent settings where the agents have different preferences, preference aggregation is a central issue. Voting is a general method for preference aggregation, but seminal results have shown that all general voting protocols are manipulable. One could try to avoid manipulation by using voting protocols where determining a beneficial manipulation is hard computationally. The complexity of manipulating realistic elections where the number of candidates is a small constant was recently studied [4], but the emphasis was on the question of *whether or not* a protocol becomes hard to manipulate for some constant number of candidates. That work, in many cases, left open the question: *How many candidates are needed to make elections hard to manipulate?* This is a crucial question when comparing the relative manipulability of different voting protocols. In this paper we answer that question for the voting protocols of the earlier study: *plurality*, *Borda*, *STV*, *Copeland*, *maximin*, *regular cup*, and *randomized cup*. We also answer that question for two voting protocols for which no results on the complexity of manipulation have been derived before: *veto* and *plurality with runoff*. It turns out that the voting protocols under study become hard to manipulate at 3 candidates, 4 candidates, 7 candidates, or never.

*Conitzer and Sandholm were funded by the National Science Foundation under CAREER Award IRI-9703122, Grant IIS-9800994, ITR IIS-0081246, and ITR IIS-0121678.

†The second author thanks Helene Fargier, Michel Lemaitre and Gerard Verfaillie for stimulating discussions about manipulation.

# 1  Introduction

Often, a group of agents has to make a common decision, yet they have different preferences about which decision is made. Thus, it is of central importance to be able to aggregate the preferences, that is, to make a socially desirable decision as to which *candidate* is chosen from a set of candidates. Such candidates could be potential presidents, joint plans, allocations of goods or resources, etc. Voting is the most general preference aggregation scheme, and has been used in several multiagent decision making problems in computer science, such as collaborative filtering (e.g. [13]) and planning among multiple automated agents (e.g. [6, 7]).

One key problem voting mechanisms are confronted with is that of *manipulation* by the voters. An agent is said to vote strategically when it does not rank the alternatives according to its true preferences, but rather so as to make the eventual outcome most favorable to itself. For example, if an agent prefers Nader to Gore to Bush, but knows that Nader has too few other supporters to win, while Gore and Bush are close to each other, the agent would be better off by declaring Gore as its top candidate. Manipulation is an undesirable phenomenon because social choice schemes are tailored to aggregate preferences in a socially desirable way, and if the agents reveal their preferences insincerely, a socially undesirable candidate may be chosen.

The issue of strategic voting has been studied extensively. A seminal negative result, the *Gibbard-Satterthwaite theorem*, shows that if there are three or more candidates, then in any nondictatorial voting scheme, there are preferences under which an agent is better off voting strategically [8, 14]. (A voting scheme is called dictatorial if one of the voters dictates the social choice no matter how the others vote.)

This problem has a strong impact on elections where voters are human. In automated group decision making where the voters are *software agents*, the manipulability of protocols is even more problematic, for at least three reasons. First, the algorithms they use to decide how to vote must be coded explicitly. Given that the voting algorithm needs to be designed only once (by an expert), and can be copied to large numbers of agents (even ones representing unsophisticated human voters), it is likely that rational strategic voting will increasingly become an issue, unmuddied by irrationality, emotions, etc. Second, software agents have more computational power and are more likely to find effective manipulations. Third, the settings where software agents are used for voting are likely to be more anonymous, and therefore the voters are likely to be less scrupulous. For example, they might not worry about the norm that manipulation, in itself, is considered inappropriate (if the voter's neighbor knows his preferences on an issue and sees him vote differently, then the neighbor will consider him a liar). As another example, anonymous voters are freed from the social pressures of voting in a particular way (e.g., voting for a school rather than a cigar factory), so the space of viable votes—and thus viable manipulations—is larger.

Nevertheless, there are several avenues for trying to avoid the possibility of manipulation. First, social choice theorists and economists have studied settings where the voters' preferences are restricted.

Under certain restrictions, such as *single-peaked preferences* or *quasilinear preferences*, nonmanipulable protocols exist (e.g. [12, 6, 7]). The weakness of this approach is that in practice the protocol designer cannot be sure that the agents' preferences fall within the restriction. If they do not, it is *impossible* to vote truthfully, so the protocol *forces* the agents to manipulate. A second approach to avoiding manipulation is randomization. For example, a dictator could be chosen at random among the voters, in which case there would be no incentive for manipulation. The randomization approach is undesirable if it introduces too much noise into the election process, and it turns out that almost complete randomization (as in the example above) is required in order to obtain a nonmanipulable protocol [9, 10]. Furthermore, randomization can introduce manipulation possibilities even when none would have existed (for the preferences that the agents happen to have) under a deterministic protocol [15].

We take a third tack toward avoiding manipulation: *ensuring that finding a beneficial manipulation is so hard computationally that it is unlikely that voters will be able to manipulate.* So, unlike in most of computer science, here high computational complexity is a desirable property. The harder it is to manipulate, the better. Especially in the context of software agents, this computational complexity is best measured with the usual tools from theoretical computer science. The approach of using computational complexity to avoid manipulation has not received much attention and many problems are still open. Bartholdi et al. prove some hardness results under the assumption that not only the number of voters but also *the number of candidates* is unbounded [2, 1]. Such results do not prove hardness of manipulation in real elections where the number of candidates is small. Conitzer and Sandholm addressed this issue by studying the complexity of manipulation when the number of candidates is a constant [4]. The emphasis in that work was on the question of *whether or not* a protocol becomes hard to manipulate for some constant number of candidates. This, in many cases, left open the question: *How many candidates are needed to make elections hard to manipulate?* This is a crucial question when comparing the relative manipulability of different voting protocols.

In this paper we answer that question for the voting protocols of the earlier study: *plurality, Borda, STV, Copeland, maximin, regular cup*, and *randomized cup*. We also answer that question for two voting protocols for which no results on the complexity of manipulation have been derived before: *veto* and *plurality with runoff*. The rest of the paper is organized as follows. In Section 2 we review standard voting protocols. In Section 3 we present definitions of manipulation. The core of the paper is Section 4 where we present our results on the complexity of manipulation. Conclusions and future research are discussed in Section 5.

## 2 Voting protocols

We now define the voting setting. Let $\mathcal{V} = \{v_1, \ldots, v_n\}$ be the finite set of *voters*. Let $\mathcal{X} = \{1, \ldots, m\}$ be the finite set of *candidates*. The *preferences* of voter $v_i$ are given by a linear order $O_i$ on $\mathcal{X}$. A *preference profile* is a vector $P = \langle O_1, \ldots, O_n \rangle$ of individual preferences.

A *voting protocol* is a function from the set of all preference profiles to the set of candidates $\mathcal{X}$.[1] The following list reviews the most common voting protocols. In the protocols that are based on scores, the candidate with the highest score wins. In each of the listed protocols (even the ones that have multiple rounds), the voters submit their preferences up front. That is, the voters are not allowed to change their preference revelations during the execution of the protocol.

- *scoring protocols.* Let $\vec{\alpha} = \langle \alpha_1, \ldots, \alpha_m \rangle$ be a vector of integers such that $\alpha_1 \geq \alpha_2 \ldots \geq \alpha_m$. For each voter, a candidate receives $\alpha_1$ points if it is ranked first by the voter, $\alpha_2$ if it is ranked second etc. The *score* $s_{\vec{\alpha}}$ of a candidate is the total number of points the candidate receives. The *Borda* protocol is the scoring protocol with $\vec{\alpha} = \langle m-1, m-2, \ldots, 0 \rangle$. The *plurality* protocol (aka. majority rule) is the scoring protocol with $\vec{\alpha} = \langle 1, 0, \ldots, 0 \rangle$. The *veto* protocol is the scoring protocol with $\vec{\alpha} = \langle 1, 1, \ldots, 1, 0 \rangle$.

- *maximin* (aka. *Simpson*). For any two distinct candidates $i$ and $j$, let $N(i,j)$ be the number of voters who prefer $i$ to $j$. The *maximin score* of $i$ is $s(i) = \min_{j \neq i} N(i,j)$.

- *Copeland.* For any two distinct candidates $i$ and $j$, let $C(i,j) = +1$ if $N(i,j) > N(j,i)$ (in this case we say that $i$ beats $j$ in their pairwise election), $C(i,j) = 0$ if $N(i,j) = N(j,i)$ and $C(i,j) = -1$ if $N(i,j) < N(j,i)$. The *Copeland score* of candidate $i$ is $s(i) = \sum_{j \neq i} C(i,j)$.

- *single transferable vote* (STV). The protocol proceeds through a series of $m-1$ rounds. At each round, the candidate with the lowest plurality score (*i.e.*, the least number of voters ranking it first among the remaining candidates) is eliminated. The winner is the last remaining candidate.

- *plurality with run-off.* In this protocol, a first round eliminates all candidates except the two with the highest plurality scores. Then votes are transferred to these (as in the *STV* protocol). After that, a second round determines the winner among these two.

- *cup* (sequential binary comparisons). The cup is defined by a balanced binary tree $T$ with one leaf per candidate, and an assignment of candidates to leaves (each leaf gets one candidate). Each non-leaf node is assigned the winner of the pairwise election of the node's children; the candidate assigned to the root wins. The *regular cup* protocol assumes that the assignment of candidates to leaves is known by the voters before they vote. In the *randomized cup* protocol [4], the assignment of candidates to leaves is chosen uniformly at random after the voters have voted. Note that the randomized cup protocol differs from all the other protocols under discussion in the sense that all the others are deterministic.

In some settings, the voters are *weighted*. A *weight function* is a mapping $w : \mathcal{V} \to \mathbb{N}^*$. When voters are weighted, the above protocols are applied by simply considering a voter of weight $k$ to be $k$ different voters. Different possible interpretations can be given to weights. They may represent the decision power of a given agent in a voting setting where not all agents are considered equal. The weight may correspond to the size of the community that the voter represents (such as the size of the state). Or, when agents vote in partisan groups (e.g., in parliament), the weights may correspond to the size of the group (each group acts as one voter).

---
[1] Some of the voting protocols require tie-breaking rules (at different stages of the execution of the protocol). These rules are usually left undefined. The results in this paper do not depend on tie-breaking rules.

## 3 Manipulating an election

In this section we define our computational problem precisely. We lead into the definition by discussing the different dimensions of the election manipulation problem:

1) *What information do the manipulators have about the nonmanipulators' votes?* In the *incomplete information* setting, the manipulators are uncertain about the nonmanipulators' votes. This uncertainty could be represented in a number of ways, for example, as a joint probability distribution over the nonmanipulators' votes. In the *complete information* setting, the manipulators know the nonmanipulators' votes exactly. We prove our results for the complete information case for the following reasons: 1a. It is a special case of any uncertainty model. Therefore, our hardness results directly imply hardness for the incomplete information setting. 1b. Via prior results of Conitzer and Sandholm [4], hardness results for manipulation by coalitions in the complete information setting also imply hardness of manipulation *by individuals* in the incomplete information setting. (We will discuss these implications in more detail in the conclusions section of this paper.) 2. Results in the complete information setting measure only the *inherent* complexity of manipulation rather than any potential complexity introduced by the model of uncertainty.

2) *Who is manipulating: an individual voter or a coalition of voters?* Both of these are important variants, but we focus on coalitional manipulation for the following reasons: 1. In elections with many voters it is very unlikely that an individual voter can affect the outcome—even with unlimited computational power. 2. For any constant number of candidates (even with an unbounded number of voters), manipulation by individuals in the complete information setting is computationally easy because the manipulator can enumerate and evaluate all its possible votes (rankings of candidates) in polynomial time [4].[2] (Manipulation by individuals in the complete information setting can be hard for some voting protocols if one allows the number of voters *and the number of candidates* to be unbounded [2, 1]). 3. Via prior results [4], hardness results for manipulation *by coalitions* in the complete information setting also imply hardness of manipulation *by individuals* in the incomplete information setting. (We will discuss these implications in more detail in the conclusions section of this paper.)

3) *Are the voters weighted or unweighted?* Both of these are important variants, but we focus on weighted voters for the following reasons: 1. In the unweighted case, for any constant number of candidates (even with an unbounded number of voters), manipulation by a coalition in the complete information setting is computationally easy because the coalition can enumerate and evaluate all its effectively different vote vectors [4]. (The number of effectively different vote vectors is polynomial due to the interchangeability of the different equiweighted voters.) 2. Via prior results [4], hardness results for manipulation by *weighted* coalitions in the complete information setting also imply hardness of evaluating the probabilities of different outcomes in the incomplete information setting with *unweighted* (but correlated) voters. (We will discuss these implications in more detail in the conclusions section of this paper.)

---

[2]This assumes that the voting protocol is easy to execute—as most protocols are (including the ones under study). However, there exist voting protocols that are NP-hard to execute [3].

4) *What is the goal of manipulation?* We study two alternative goals: trying to make a given candidate win (we call this *constructive* manipulation), and trying to make a given candidate *not* win (we call this *destructive* manipulation). Besides these goals being elegantly crisp, there are fundamental theoretical reasons to focus on these goals:

First, hardness results for these goals imply hardness of manipulation under any game-theoretic notion of manipulation, because our manipulation goals are always special cases. (This holds both for deterministic and randomized voting protocols.) At one extreme, consider the setting where there is one candidate that would give utility 1 to each of the manipulators, and all other candidates would give utility 0 to each of the manipulators. In this case the only sensible game-theoretic goal for the manipulators is to make the preferred candidate win. This is exactly our notion of constructive manipulation. At the other extreme, consider the setting where there is one candidate that would give utility 0 to each of the manipulators, and all other candidates would give utility 1 to each of the manipulators. In this case the only sensible game-theoretic goal for the manipulators is to make the hated candidate not win. This is exactly our notion of destructive manipulation.

Second, at least for deterministic voting protocols in the complete information setting, the easiness results transfer from constructive manipulation to any game-theoretic definitions of manipulation that would come down to determining whether the manipulators can make some candidate from a subset of candidates win. For example, one can consider a manipulation successful if it causes some candidate to win that is preferred by each one of the manipulators to the candidate who would win if the manipulators voted truthfully. As another example, one can consider a manipulation successful if it causes some candidate to win that gives a higher sum of utilities to the manipulators than the candidate who would win if the manipulators voted truthfully. (This definition is especially pertinent if the manipulators can use side payments or some other form of restitution to divide the gains among themselves.) Now, we can solve the problem of determining whether some candidate in a given subset can be made to win simply by determining, for each candidate in the subset in turn, whether that candidate can win. So the complexity exceeds that of constructive manipulation by at most a factor equal to the number of candidates (i.e., a constant).

Third, the complexity of destructive manipulation is directly related to the complexity of determining whether enough votes have been elicited to determine the outcome of the election. Specifically, enough votes have been elicited if there is no way to make the conjectured winner not win by casting the yet unknown votes [5].

In summary, we focus on coalitional weighted manipulation (*cw-manipulation*), in the complete information setting. We study both constructive and destructive manipulation. Formally:

**Definition 1** CONSTRUCTIVE COALITIONAL WEIGHTED (CW)-MANIPULATION. *We are given a set of weighted votes $S$ (the nonmanipulators' votes), the weights for a set of votes $T$ which are still open (the manipulators' votes), and a preferred candidate $p$. For deterministic protocols, we are asked whether there is a way to cast the votes in $T$ so that $p$ wins the election. For randomized protocols, we are additionally given a distribution over instantiations of the voting protocol, and a number $r$, where $0 \leq r \leq 1$. We are asked whether there is a way to cast the votes in $T$ so that $p$ wins with probability greater than $r$.*

**Definition 2** DESTRUCTIVE COALITIONAL WEIGHTED (CW)-MANIPULATION. *We are given a set of weighted votes $S$ (the nonmanipulators' votes), the weights for a set of votes $T$ which are still open (the manipulators' votes), and a hated candidate $h$. For deterministic protocols, we are asked whether there is a way to cast the votes in $T$ so that $h$ does* not *win the election. For randomized protocols, we are additionally given a distribution over instantiations of the voting protocol, and a number $r$, where $0 \leq r \leq 1$. We are asked whether there is a way to cast the votes in $T$ so that $h$ wins with probability* less *than $r$.*

## 4  Complexity of manipulation

Now, how many candidates are needed to make an election computationally hard to manipulate? The answer depends on the voting protocol, and whether we are interested in constructive or destructive manipulation. For example, the plurality protocol is easy to manipulate constructively and destructively for any number of candidates [4]. The Borda protocol becomes hard to manipulate constructively with 3 candidates already, but is easy to manipulate destructively for any number of candidates [4].

The complexity of manipulation with a finite number of candidates has been studied by Conitzer and Sandholm [4] for many of the voting protocols discussed in this paper. However, their focus was on the question of whether or not a protocol is hard to manipulate for *some* finite number of candidates. For many of the protocols, they did not determine the exact number of candidates at which the polynomial to NP-complete transition occurs. This number is important for evaluating the relative manipulability of different voting protocols (the lower this number, the less manipulable the protocol). In this paper, we determine the exact number of candidates where this transition occurs. We also determine this number for two protocols not addressed by Conitzer and Sandholm: veto and plurality with runoff.

When there are only two candidates, all the protocols are equivalent to the plurality protocol, and hence both types of manipulation (constructive and destructive) are in P for all of the protocols. The following theorem summarizes the results of Conizer and Sandholm [4] about the complexity of manipulation as the number of candidates exceeds two:

**Theorem 1** *[4]*

- *For the* Borda *and* STV *protocols,* CONSTRUCTIVE CW-MANIPULATION *is* NP-*complete for* $\geq 3$ *candidates;*

- *For the* Copeland *and* maximin *protocols,* CONSTRUCTIVE CW-MANIPULATION *is* NP-*complete for* $\geq 4$ *candidates;*

- *For the* regular cup *and* plurality *protocols,* CONSTRUCTIVE CW-MANIPULATION *and* DESTRUCTIVE CW-MANIPULATION *are in* P *regardless of the number of candidates;*

- *For the* randomized cup *protocol,* CONSTRUCTIVE CW-MANIPULATION *is* NP-*complete for* $\geq 7$ *candidates;*

- *For the* Borda, Copeland *and* maximin *protocols,* DESTRUCTIVE CW-MANIPULATION *is in* P *regardless of the number of candidates;*

- *For the* STV *protocol,* DESTRUCTIVE CW-MANIPULATION *is* NP-*complete for* $\geq 4$ *candidates.*

In the next subsection, we present our new results on the complexity of constructive manipulation. In the subsection after that, we lay out our new results on the complexity of destructive manipulation.

## 4.1 New results on the complexity of *constructive* manipulation

We present our hardness results first, followed by our easiness results.

### 4.1.1 Hardness results

In this section we show hardness results for the two voting protocols for which no hardness results were known before. In many of the proofs of NP-hardness, we use a reduction from the PARTITION problem, which is NP-complete [11]:

**Definition 3** PARTITION. *We are given a set of integers $\{k_i\}_{1 \leq i \leq t}$ (possibly with multiplicities) summing to $2K$, and are asked whether a subset of these integers sums to $K$.*

Now we are ready to show our hardness results about manipulating elections.

**Theorem 2** *For the* veto *protocol,* CONSTRUCTIVE CW-MANIPULATION *is* NP-*complete for 3 candidates and more.*

**Proof.** Showing the problem is in NP is easy. To show it is NP-hard, we reduce an arbitrary PARTITION instance to the following constructive cw-manipulation instance. The 3 candidates are $a$, $b$ and $p$. In $S$ there are $2K - 1$ voters voting $(a, b, p)$ (vetoing $p$). In $T$, for every $k_i$ there is a vote of weight $2k_i$. We show the instances are equivalent.

If a partition of the $k_i$ exists, let the votes in $T$ corresponding to one half of the partition vote $(p, a, b)$ (vetoing $b$), and let the other ones vote $(p, b, a)$ (vetoing $a$). Then $p$ has $4K$ points (it is vetoed by $2K - 1$ of the vote weight), whereas $a$ and $b$ each have only $4K - 1$ points (they are each vetoed by $2K$ of the vote weight). So there exists a manipulation.

On the other hand, if a manipulation exists, let $A$ be the set of $k_i$ corresponding to votes in the manipulation vetoing $a$, and let $B$ be the set of $k_i$ corresponding to votes in the manipulation vetoing $b$. Because $p$ is vetoed by at least $2K - 1$ of the vote weight, and none of the votes in $S$ veto $a$, it follows that at least $2K - 1$ of the vote weight in $T$ vetoes $a$, that is, $\sum_{k_i \in A} 2k_i \geq 2K - 1$, or $\sum_{k_i \in A} k_i \geq K - \frac{1}{2}$. Because the $k_i$ are integers, it follows that $\sum_{k_i \in A} k_i \geq K$. Similarly, $\sum_{k_i \in B} k_i \geq K$. Since $A$ and $B$ are disjoint, it follows that $\sum_{k_i \in A} k_i = \sum_{k_i \in B} k_i = K$. So there exists a partition. ∎

**Theorem 3** *For the* plurality with runoff *protocol,* CONSTRUCTIVE CW-MANIPULATION *is* NP-*complete for 3 candidates and more.*

**Proof.** Showing the problem is in NP is easy. To show it is NP-hard, we observe that with 3 candidates, *plurality wit runoff* coincides with $STV$, and CONSTRUCTIVE MANIPULATION for $STV$ with 3 candidates is NP-hard, as we summarized in Theorem 1. ∎

### 4.1.2 Easiness results

We now solve several questions that were left unanswered in [4]. As was shown, CW-CONSTRUCTIVE MANIPULATION for the Copeland and maximin protocols is NP-complete for 4 candidates and more, and in P for 2 candidates. But what about the case of 3 candidates? This problem is even more intriguing for the *randomized cup*: CW-CONSTRUCTIVE MANIPULATION is NP-complete for 7 candidates and more, and in P for 2 candidates. But what if the number of candidates is in the 3–6 range? In this section we address these three questions in order.

In each case, we prove easiness by demonstrating that if there is a successful manipulation, there also exists one where all the manipulators vote the same way. Because the number of candidates is constant, the number of different orderings of the candidates is constant. Therefore all such ways of voting can be easily enumerated (and each way of voting is easy to evaluate in these protocols).

**Theorem 4** *If the* Copeland *protocol with 3 candidates has a* CONSTRUCTIVE CW-MANIPULATION*, then it has a* CONSTRUCTIVE CW-MANIPULATION *where all of the manipulators vote identically. Therefore,* CONSTRUCTIVE CW-MANIPULATION *is in* P.

**Proof.** Let the 3 candidates be $p$, $a$, and $b$. We are given the nonmanipulators' votes $S$, and the weights for the manipulators' votes $T$. Let the total vote weight in $T$ be $K$.

For a set of weighted votes $V$ and two candidates $x$, $y$, we denote by $N_V(x, y)$ the cumulated weights of the votes in $V$ ranking $x$ prior to $y$, and we let $D_V(x, y) = N_V(x, y) - N_V(y, x)$. Let us consider the following four cases which cover all possible situations:

*Case 1*: $K > D_S(a, p)$ and $K > D_S(b, p)$.

In this case, *any* configuration of votes for $T$ such that $p$ is ranked first for all votes makes $p$ win the election.

*Case 2*: $K > D_S(a, p)$ and $K = D_S(b, p)$.

It can easily be shown that it is harmless to assume that all votes in $T$ rank $p$ first. Therefore, what remains to be done in order to have $p$ win is to find who in $T$ should vote $(p, a, b)$ and who should vote $(p, b, a)$. What we know so far (before knowing how the votes in $T$ will split between these two profiles) is: (1) $D_{S \cup T}(p, a) = K - D_S(p, a) > 0$ and (2) $D_{S \cup T}(p, b) = K - D_S(p, b) = 0$. (1) makes $p$ get $+1$ and $a$ get $-1$ while (2) makes both $p$ and $b$ get 0. Therefore, the partial Copeland scores (not taking account of the $a$ vs. $b$ pairwise election), are $+1$ for $p$ (which will not change after taking account of the $a - b$ pairwise election), $-1$ for $a$ and 0 for $b$; hence, the only way for $p$ to win (with certainty) is to avoid $b$ getting a point in the pairwise election against $a$, i.e., to ensure that $D_{S \cup T}(a, b) \geq 0$. It can easily be shown that this is possible if and only if $K \geq D_S(b, a)$. Therefore, we have found that there exists a successful manipulation for $p$ iff $K \geq D_S(b, a)$, and in this case a successful manipulation is the one where all voters in the coalition vote $(p, a, b)$.

*Case 3*: $K = D_S(a, p)$ and $K > D_S(b, p)$.

This is similar to Case 2, switching the roles of $a$ and $b$; the condition then is $K \geq D_S(a, b)$ and the successful manipulation is the one where all vote $(p, b, a)$.

*Case 4*: $K < D_S(a, p)$ or $K < D_S(b, p)$ or $(K \leq D_S(a, p)$ and $K \leq D_S(b, p))$.

Here, whatever the votes in $T$, the Copeland score of $p$ is smaller than or equal to 0 and therefore $p$ cannot be guaranteed to win, so there is no successful manipulation. Thus, in every case, either there is no successful manipulation, or there is a successful manipulation where all manipulators vote identically. ∎

**Theorem 5** *If the* maximin *protocol with 3 candidates has a* CONSTRUCTIVE CW-MANIPULATION, *then it has a* CONSTRUCTIVE CW-MANIPULATION *where all of the manipulators vote identically. Therefore,* CONSTRUCTIVE CW-MANIPULATION *is in* P.

**Proof.** Let the 3 candidates be $p$, $a$, and $b$. We are given the nonmanipulators' votes $S$, and the weights for the manipulators' votes $T$. Let the total vote weight in $T$ be $K$. Again, it is easy to show that all the manipulators can rank $p$ first without harm.

Let us denote by $P_{K1,K2}$ a vote configuration for $T$ such that a subset $T_1$ of $T$, whose cumulated weight is $K_1$, votes $(p, a, b)$ and $T_2 = T \setminus T_1$, whose cumulated weight is $K_2$ (with $K_1 + K_2 = K$), votes $(p, b, a)$. Now all that remains to show is the following: if $p$ wins with the votes in $T$ being $P_{K1,K2}$ then either $p$ wins with the votes in $T$ being $P_{K,0}$ or $p$ wins with the votes in $T$ being $P_{0,K}$. Let us consider these two cases for the outcome of the whole election (including the votes in $T$):

*Case 1*: the uniquely worst pairwise election for $a$ is against $b$, and the uniquely worst pairwise election for $b$ is against $a$. One of $a$ and $b$ must have got at least half the vote weight in the pairwise election against the other (say, WLOG, $a$) and therefore have a maximin score of at least half the vote weight. Since $a$ did even better against $p$, $p$ received less than half the vote weight in their pairwise election and therefore $p$ does not win.

*Case 2*: One of $a$ and $b$ (say, WLOG, $a$) does at least as badly against $p$ as against the other (so, $a$'s worst opponent is $p$). Then all the voters in the coalition might as well vote $(p, a, b)$, because this will change neither $a$'s score nor $p$'s score, and might decrease (but not increase) $b$'s score. ∎

**Theorem 6** *If the randomized cup protocol with 6 candidates has a* CONSTRUCTIVE CW-MANIPULATION, *then it has a* CONSTRUCTIVE CW-MANIPULATION *where all of the manipulators vote identically. (This holds regardless of which balanced tree is chosen.) Therefore,* CONSTRUCTIVE CW-MANIPULATION *is in* P.

**Proof.** Omitted for reasons of space constraint. ∎

## 4.2 New results on the complexity of *destructive* manipulation

In this section we present our new results on the complexity of destructive manipulation. We first lay out the hardness results, and then the easiness results.

### 4.2.1 Hardness results

Among the protocols whose complexity has been studied with respect to destructive manipulation, the only unanswered question is the complexity of DESTRUCTIVE CW-MANIPULATION of the $STV$ protocol with 3 candidates. We now determine that complexity.

**Theorem 7** *For the* STV *protocol with 3 candidates,* DESTRUCTIVE CW-MANIPULATION *is* NP-*complete*.

**Proof.** Showing the problem is in NP is easy. To show it is NP-hard, we reduce an arbitrary PARTITION instance to the following DESTRUCTIVE CW-MANIPULATION instance. The 3 candidates are $a$, $b$ and $h$. In $S$ there are $6K$ voters voting $(a, h, b)$, $6K$ voters voting $(b, h, a)$, and $8K - 1$ voters voting $(h, a, b)$. In $T$, for every $k_i$ there is a vote of weight $2k_i$. We show the instances are equivalent.

We first observe that $h$ will not win if and only if it gets eliminated in the first round: for if it survives the first round, either $a$ or $b$ gets eliminated in the first round. Hence either all the votes in $S$ that ranked $a$ at the top or all those that ranked $b$ at the top will transfer to $h$, leaving $h$ with at least $14K - 1$ votes in the final round out of a total of $24K - 1$, so that $h$ is guaranteed to win the final round.

Now, if a partition of the $k_i$ exists, let the votes in $T$ corresponding to one half of the partition vote $(a, b, h)$, and let the other ones vote $(b, a, h)$. Then in the first round, $a$ and $b$ each have $8K$ votes, and $h$ only has $8K - 1$ votes, so that $h$ gets eliminated. So there exists a manipulation.

On the other hand, if a manipulation exists, we know by the above that with this manipulation, $h$ is eliminated in the first round. Hence at least $2K - 1$ of the vote weight in $T$ ranks $a$ at the top, and at least $2K - 1$ of the vote weight in $T$ ranks $b$ at the top. Let $A$ be the set of all the $k_i$ corresponding to votes in $T$ ranking $a$ at the top; then $\sum_{k_i \in A} k_i \geq K - \frac{1}{2}$, and since the $k_i$ are integers this implies

$\sum\limits_{k_i \in A} k_i \geq K$. If we let $B$ be the set of all the $k_i$ corresponding to votes in $T$ ranking $b$ at the top, then similarly, $\sum\limits_{k_i \in B} k_i \geq K$. Since $A$ and $B$ are disjoint, it follows that $\sum\limits_{k_i \in A} k_i = \sum\limits_{k_i \in B} k_i = K$. So there exists a partition. ∎

This result also allows us to establish the hardness of destructive manipulation in the plurality with runoff protocol:

**Theorem 8** *For the* plurality with runoff *protocol with 3 candidates,* DESTRUCTIVE CW-MANIPULATION *is* NP-*complete .*

**Proof.** Showing the problem is in NP is easy. To show it is NP-hard, we observe that with 3 candidates, *plurality wit runoff* coincides with $STV$, and DESTRUCTIVE MANIPULATION for $STV$ with 3 candidates is NP-hard , as we proved in Theorem 7. ∎

#### 4.2.2 Easiness result

While plurality with runoff is hard to manipulate destructively, the other protocol that was not studied before turns out to be easy to manipulate destructively:

**Theorem 9** *For the* veto *protocol,* DESTRUCTIVE CW-MANIPULATION *is in* P *for any number of candidates.*

**Proof.** Conitzer and Sandholm showed that for protocols in which each candidate gets a numerical score (and the candidate with the highest score wins), and which are *monotonic*, DESTRUCTIVE CW-MANIPULATION is in P for any number of candidates. [4] (A score-based protocol is *monotonic* if ranking a candidate higher never decreases that candidate's score.) It is immediate from our definition of the veto protocol that it is a monotonic score-based protocol. ∎

## 5 Conclusions and future research

In multiagent settings where the agents have different preferences, preference aggregation is a central issue. Voting is a general method for preference aggregation, but seminal results have shown that all general voting protocols are manipulable. One could try to avoid manipulation by using voting protocols where determining a beneficial manipulation is hard. Especially among computational agents, it is reasonable to measure this hardness by computational complexity. Most earlier research on this topic assumed that the number of voters and candidates is unbounded [2, 1]. Such hardness results lose relevance when the number of candidates is small, because manipulation algorithms that are exponential only in the number of candidates might be available. The complexity of manipulating realistic elections where the number of candidates is a small constant was recently studied [4], but the emphasis was on the question of *whether or not* a protocol becomes hard to manipulate for some constant number of candidates. That work, in many cases, left open the question: *How many candidates are needed to make*

*elections hard to manipulate?* This is a crucial question when comparing the relative manipulability of different voting protocols.

In this paper we answered that question for the voting protocols of the earlier study. We also answered the question for two voting protocols for which no results on the complexity of manipulation had been derived before: veto and plurality with runoff.

The following tables summarize the complexity of constructive and destructive manipulation, respectively, as the state of knowledge stands after this paper. The nontrivial new results of this paper are marked by an asterisk (*).

| Number of candidates | 2 | 3 | 4,5,6 | ≥ 7 |
|---|---|---|---|---|
| *Borda* | P | NP-c | NP-c | NP-c |
| *veto* | P | NP-c* | NP-c* | NP-c* |
| *STV* | P | NP-c | NP-c | NP-c |
| *plurality with runoff* | P | NP-c* | NP-c* | NP-c* |
| *Copeland* | P | P* | NP-c | NP-c |
| *maximin* | P | P* | NP-c | NP-c |
| *randomized cup* | P | P* | P* | NP-c |
| *regular cup* | P | P | P | P |
| *plurality* | P | P | P | P |

Complexity of CONSTRUCTIVE CW-MANIPULATION

| Number of candidates | 2 | ≥ 3 |
|---|---|---|
| *STV* | P | NP-c* |
| *plurality with runoff* | P | NP-c* |
| *randomized cup* | P | ? |
| *Borda* | P | P |
| *veto* | P | P* |
| *Copeland* | P | P |
| *maximin* | P | P |
| *regular cup* | P | P |
| *plurality* | P | P |

Complexity of DESTRUCTIVE CW-MANIPULATION

As in the earlier paper, we studied manipulation by coalitions where the voters are weighted. In this paper we clearly justified why that is a key setting to study, and also layed out the key reasons why our definitions of manipulation are central. Our hardness results also imply hardness for manipulation *by individuals* when the manipulator is uncertain about the nonmanipulators' votes.[3]

Future research includes determining the complexity of destructively manipulating the randomized cup protocol with 3 or more candidates, and designing new voting protocols that are hard to manipulate according to our measure. Finally, NP-hardness is a worst-case measure, and it would be desirable to prove that some of the voting protocols are hard to manipulate *on average*, or to design entirely new voting protocols that are.

---

[3]This implication is based on an earlier general theorem of how coalitional manipulation results derived in the complete information setting directly translate to individual manipulation in the incomplete information setting [4]. Our hardness results also translate to manipulations by individuals in the *unweighted* setting—at least as long as the votes of the nonmanipulators are allowed to be correlated.